\documentclass[prl,twocolumn,showpacs,showpacs,superscriptaddress]{revtex4}

\usepackage{graphicx}
\usepackage{color}

\newcommand{\tr}{\mbox{tr}}

\begin{document}

\title{Shot noise and magnetism of Pt atomic chains: accumulation of points at the boundary }
\author{Manohar Kumar}
\affiliation{Kamerlingh Onnes Laboratorium, Universiteit Leiden,
Postbus 9504, 2300 RA Leiden, The Netherlands}
\affiliation{Cavendish Laboratory, J.J. Thomson Avenue, Cambridge CB3 0HE, UK}
\author{Oren Tal}
\affiliation{Kamerlingh Onnes Laboratorium, Universiteit Leiden,
Postbus 9504, 2300 RA Leiden, The Netherlands}
\affiliation{Department of Chemical Physics, Weizmann Institute of
Science, Rehovot 76100, Israel}
\author{Roel H.M. Smit}
\affiliation{Kamerlingh Onnes Laboratorium, Universiteit Leiden,
Postbus 9504, 2300 RA Leiden, The Netherlands}
\author{Alexander Smogunov}
\affiliation{CEA, IRAMIS, SPCSI, F-91191 Gif-sur-Yvette Cedex, France}
\author{Erio Tosatti}
\affiliation{SISSA, Via Bonomea 265, Trieste 34136, Italy}
\affiliation{CNR-IOM Democritos, Via Bonomea 265, Trieste 34136, Italy}
\affiliation{ICTP, Strada Costiera 11, Trieste 34151, Italy}
\author{Jan M. van Ruitenbeek$^{*}$}
\email[Corresponding author:  ]{ruitenbeek@physics.leidenuniv.nl}
\affiliation{Kamerlingh Onnes Laboratorium, Universiteit Leiden,
Postbus 9504, 2300 RA Leiden, The Netherlands}


\begin{abstract}
Pt is known to show spontaneous formation of monatomic chains 
upon breaking a metallic contact. From model calculations these chains
are expected to be spin polarized.
However, direct experimental evidence for or against magnetism is lacking. 
Here, we investigate shot noise as a potential source of 
information on the magnetic state of Pt atomic chains. 
We observe a remarkable structure in the distribution of measured shot noise levels,
where the data appear to be confined to the region of non-magnetic states. 
While this suggests a non-magnetic ground state for the Pt atomic chains, 
from calculations we find that the magnetism in Pt chains is due to `actor' electron 
channels, which contribute very little to ballistic conductance and noise. On the other hand, 
there are weakly polarized `spectator' channels, which carry most of the current and are only slightly modified by the magnetic state. 
\end{abstract}
\maketitle

\section{Introduction}
At nanometer scales one often finds unexpected behavior of
matter. A particularly appealing example is the spontaneous
formation of chains of metal atoms upon breaking a metallic
contact \cite{yanson98,ohnishi98}. Pt is a metal with a modestly
Stoner-enhanced magnetic susceptibility, indicating proximity to a
ferromagnetic state. A transition to ferromagnetism can be induced
by reducing dimensions, as evidenced by recent work on Pt clusters
\cite{liu06}. In addition, the cylindrical symmetry for a monatomic chain leads to
partial de-quenching of the orbital angular momentum, which boosts spin magnetism
with an orientation parallel to the chain axis \cite{smogunov08a}. 
For these reasons, the ferromagnetic order predicted from model calculations for atomic chains 
\cite{delin03,fernandez05,smogunov08,smogunov08a,thiess09} is not unexpected. 
Experimentally on the other hand, it is very hard to design a
probe that can directly measure the magnetism of atomic chains. 
Here we investigate shot noise, the intrinsic quantum noise due to
the discrete character of the electronic charge, and demonstate
that it is in principle capable of revealing information on the magnetic state of Pt atomic chains.
The data appear to point at a non-magnetic ground state, as we have argued in a preliminary report \cite{preprint}. 
However, we show by model calculations that  
the ballistic conductance in a magnetic Pt chain is dominated by weakly polarized, $s$-derived  channels,
whose transmission and noise are largely spin independent. Strong spin polarization  
belongs to a subset of $d$-derived bands that are poorly involved in conductance, and rather spectators of 
the electron transmission process.  
In the experiment we observe a remarkable accumulation of points at the non-magnetic boundary, which 
can be understood as resulting from these two classes of conductance channels.

Recently Wang and Wang \cite{wang11} have also reported theoretical work on conductance and shot noise 
in Pt chain contacts, at a large strain of $d=2.8$\AA. 
Also in that work the contacts were predicted to develop magnetism, but large values for shot noise were reported, with points far away from the nonmagnetic boundary. However, it is important to note that spin-orbit effects -- which are very large in Pt --  
were ignored, as well as local Coulomb interactions, resulting in band structures and conductance channels 
that are very different from those presented in the present paper.

Shot noise was first discussed for vacuum diodes by Schottky
\cite{schottky18}, who showed that this current noise is
independent of frequency (white noise) up to very high
frequencies, and its power spectrum has a value of $S_I=
2e\overline{I}$, with $e$ the absolute value of the electron
charge, and $\overline{I}$ the average current. In nanoscale
conductors, for which the system size is much smaller than the
electron scattering length, this noise can be understood as
partition noise. In these systems the number of transmission
channels available for electrons to cross a conductor is limited
and the transmission through each one of the channels is set by
the properties of the conductor. When the transmission probability
is smaller than 1 the conductor can be viewed as an effective
bottleneck causing a random sequence of electron backscattering
events, which is observed as current fluctuations or noise. The
theory has been elaborated by several groups and has been
thoroughly reviewed by Blanter and B{\"u}ttiker \cite{blanter00}.
For a nanoscale conductor with $N$ conductance channels, each
characterized by a transmission probability $\tau_{n}$, the
current noise power at an applied bias voltage $V$ is given by,
\begin{equation}
S_I = 2eV \coth \left( \frac{eV}{2k_{\rm B} T} \right)
\frac{e^2}{h} \sum_{n=1}^N \tau_{n} (1-\tau_{n})
+  4k_{\rm B} T \frac{e^2}{h}  \sum_{n=1}^N \tau_n ^2.
\label{eq:coth}
\end{equation}
\noindent where $k_{\rm B}$ is Boltzmann's constant, and $T$ is
the temperature of the nanoscale conductor. Anticipating spin
splitting of the conductance channels we treat conductance
channels for each spin direction separately. In equilibrium (at
$V=0$), Eq.~(\ref{eq:coth}) reduces to the Johnson-Nyquist
thermal noise, $4k_{\rm B} TG$, describing the current
fluctuations that are driven only by the thermal motion of
electrons. $G=(e^2/h)\sum \tau_n$ is the conductance. Again, in
the expression for $G$ we take the conductance quantum as $e^2/h$
and sum over spin states. In the low-temperature limit, $k_{\rm B}
T \ll eV$, equation~(\ref{eq:coth}) reduces to $S_I=
2e\overline{I}F$, where the Fano factor $F$ measures the quantum
suppression of Schottky's classical result ($S_I=
2e\overline{I}$),
\begin{equation}
F=\frac{\sum_n \tau_n (1-\tau_n)}{\sum_n \tau_n}.
\end{equation}
From this analysis it is apparent that one may obtain information
on the transmission probabilities of the conductance channels by
measurement of the noise power, and in favorable cases it is even
possible to determine the number of conductance channels
\cite{brom99}. The Fano factor reduces to zero when all
conductance channels are either fully blocked ($\tau_n = 0$), or
fully open ($\tau_n =1$). For a nanowire with a given conductance
$G=(e^2/h)\sum \tau_n$ the noise has a lower bound that is
obtained by taking all open channels to have perfect transmission,
except for one that takes the remaining fraction of the
conductance. This minimum will sensitively depend on whether the
spin channels are restricted to be degenerate. It is this property
that we attempt at exploiting when investigating the magnetic state of Pt atomic
chains.

\begin{figure}[t!]
\includegraphics[width=8cm]{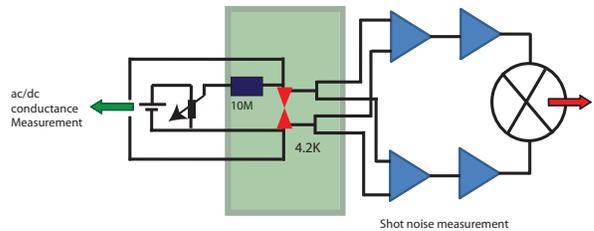}
\caption{(Color online) Schematics of the circuit used for
measuring conductance and noise on an atomic junction formed by
the MCBJ technique. Noise is measured by using two sets of
low-noise amplifiers, with a total amplification factor of $10^5$,
and by taking the cross spectrum of the two channels in a
frequency range between 250Hz and 100kHz. After averaging of
$10^4$ spectra the uncorrelated noise of the preamplifiers is
strongly suppressed.} \label{fig:circuit}
\end{figure}

\section{Shot noise measurements}

Platinum atomic junctions were formed at liquid helium
temperatures using mechanically controllable break junctions
(MCBJ, for more details see Refs.~\cite{djukic06,agrait03,suppl}).
The electronic circuit for the measurements is shown schematically
in Fig~\ref{fig:circuit}. The Pt contacts were first characterized
by recording conductance histograms \cite{smit01} as presented in Ref.~\cite{suppl} Fig.~S1. The
conductance histograms show a first peak at a conductance of about
$1.5~(2e^2/h$) with very few conductance counts below
$1~(2e^2/h)$, as expected for clean Pt point contacts: Pt
being an $s$-$d$ metal has up to 12 conductance channels due to the
six $s$ and $d$ orbitals, and spin. Each of the channels has a finite
transmission probability and they sum up to a total of about
$1.5~(2e^2/h$), in agreement with calculations
\cite{nielsen02,vega04,fernandez05, pauly06, smogunov08}. The strong peak at
$1.5~(2e^2/h$) reflects the frequent formation of atomic
chains in the contact. Chain formation can be demonstrated more
explicitly by recording histograms of the length of the
conductance plateaux (see Ref.~\cite{suppl} Fig.~S2)
with conductance values in the range of the first
conductance peak, between 1.2 and 2 times ($2e^2/h$)
\cite{yanson98,smit01,untiedt02}.

\begin{figure}[t!]
\includegraphics[width=6cm]{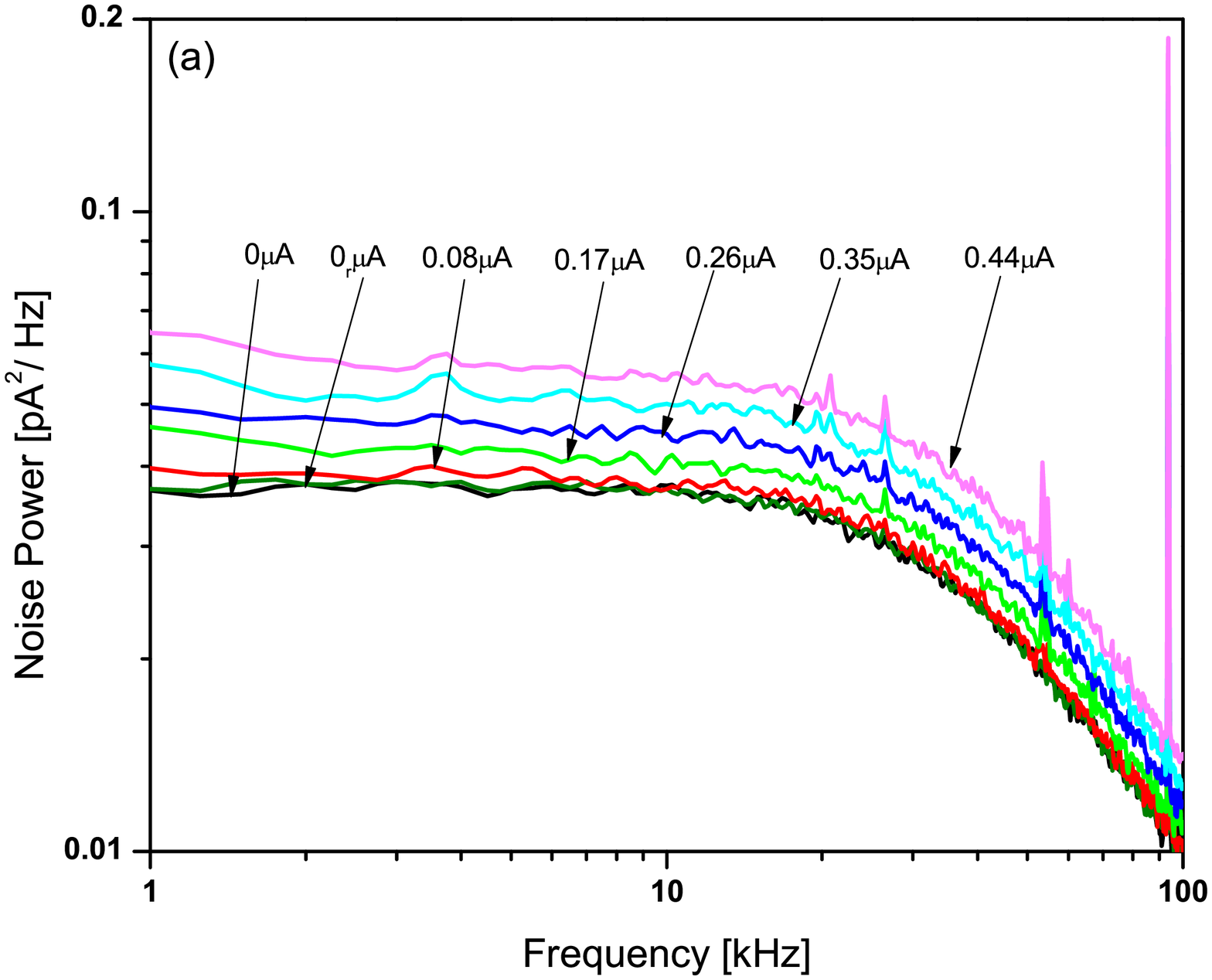}
\includegraphics[width=7cm]{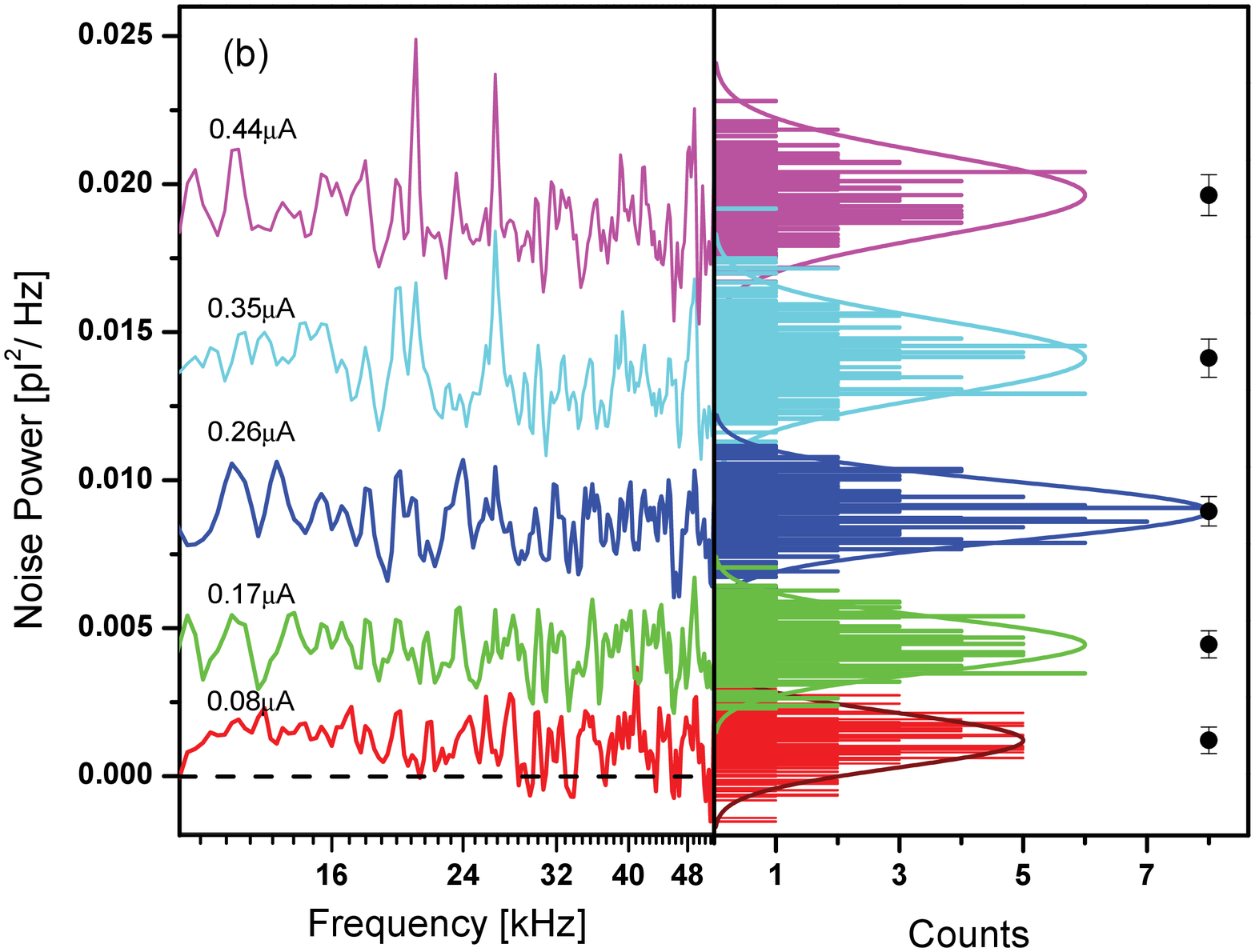}
\caption{ Example of the noise power data analysis. (a) Noise
power spectra for a Pt atomic chain of $\sim 3$ atoms in length
having a conductance $G=1.52~(2e^2/h)$ and a Fano factor
F=0.192. The peaks are due to spurious signals that could not be
fully shielded. (b) Same data after subtracting the thermal noise
and correcting for the roll-off. The spurious signals are
effectively removed by the subtraction procedure. }
\label{fig:raw_data}
\end{figure}

After this preliminary characterization of the junction an atomic
chain was made by pulling, starting from a large contact until the
conductance was seen to drop to a value near $1.8~(2e^2/h$). Measurements
of conductance and noise were taken at several points of
subsequent stretching starting from here. The corresponding piezo
voltages were recorded in order to identify the length in terms of
the mean number of atoms forming the chain. The zero-bias
differential conductance, $dI/dV$, was recorded, which is needed
in combination with the noise for the analysis of the conductance
channels. The accuracy of the ac conductance measurement is better
than 1\%, as verified by tests on standard resistors.
Typically, the sequence requires about 20min of measurement time on a junction.
 Figure~\ref{fig:raw_data} shows an example of
noise spectra taken in the window from 1 to 100~kHz for a series
of current settings, and illustrates how the noise power is
obtained from the data. First, the thermal noise is recorded at
zero bias, and after taking noise spectra at several bias settings
the zero bias noise is recorded once more (labeled as $0_r
\rm{\mu A}$)  in order to verify that the junction has remained
stable. The low-frequency upturn at larger currents is due to
$1/f$-like noise. At high frequencies there is a roll-off due to the
transfer characteristics of the circuit, with time constant RC. The thermal noise level
corresponds to a temperature of 6.3 K, which agrees within the
accuracy of the temperature measurement with a reading of 6.1 K,
as obtained from a ruthenium oxide 10k$\Omega$ resistance thermometer. For
several junction settings conductance measurements were repeated
after the shot noise bias sequence in order to detect possible
changes in the conductance. Typical changes observed were smaller
than 2\%. Figure~\ref{fig:raw_data}(b) shows that the spectra become
white above 10kHz after correction for the roll-off with a single
RC time constant. The thermal noise (at zero bias) is subtracted,
which explains the negative values in the data fluctuations for
the lowest currents. The data points are projected in the form of
a histogram, shown at the right, and the level of white noise is
obtained from the center of the histogram for each voltage bias.
The bullets and error bars at the right indicate the position and
accuracy of the noise power as determined from a gaussian fit to
the histograms.

Since shot and thermal noises are of comparable magnitude in
these experiments it is useful to represent the data such that the
expected dependence on the applied bias in Eq.~(\ref{eq:coth}) is
apparent. The voltage dependence in Eq.~(\ref{eq:coth}) can be
lumped into a single variable $X$ that we take to be $X=x\coth x$,
with $x=eV/2k_{\rm B}T$. The reduced excess noise is then defined
as,
\begin{equation}
Y=\frac{S_I (V) -S_I (0)}{S_I (0)},
\end{equation}
where $S_I (V)$ is the noise at finite bias, and $S_I (0)$ is the
thermal noise, at zero bias. The reduced excess noise is now
expected to depend linearly on the control parameter, $Y=(X-1)F$,
from which the Fano factor $F$ can be easily obtained.

\begin{figure}[b!]
\includegraphics[width=8cm]{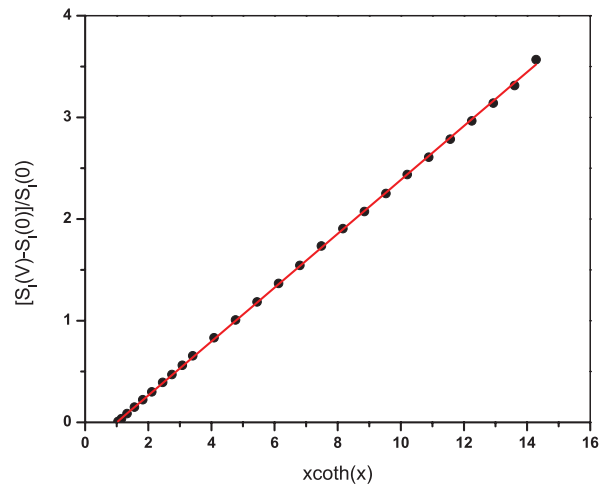}
\caption{Reduced excess noise $Y=(S_I (V) -S_I(0))/S_I (0)$ for a
Pt atomic chain. The excess noise is given as a function of
$X=x\coth (x)=(eV/2k_{\rm B} T)\coth (eV/2k_{\rm B} T)$, for a
chain having a conductance of $G=1.425\pm 0.01(2e^2/h)$ at a
length of about two atoms in the chain.  } \label{fig:reduced-plot}
\end{figure}

Figure~\ref{fig:reduced-plot} shows a series of measurements on a
Pt atomic chain with a conductance of $G=1.425\pm 0.01(2e^2/h)$ at
a short length of 2 atoms in the chain, for 26 settings of the
bias voltage in the range from  0 to 16.6~mV (0 to 1.83$\rm{\mu
A}$). The slope of the plot gives a Fano factor $F=0.269\pm
0.009$. The accuracy for each of the points is 3\%, as obtained by
a fit to the power spectrum after correction for the roll-off as
in Fig.~\ref{fig:raw_data}. The measurement required about 50
minutes, illustrating the long-term stability of the atomic
chains. Figure~\ref{fig:reduced-plot} shows a very nice agreement with the expected
dependence, and the scatter around the linear slope is within the
data point accuracy.

\begin{figure}[t!]
\includegraphics[width=8.7cm]{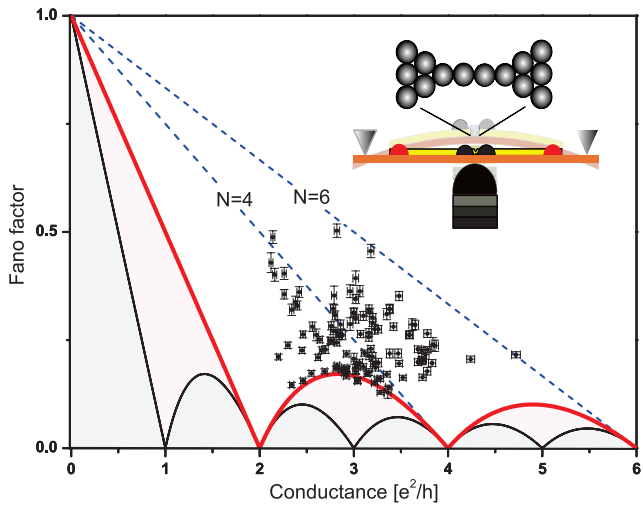}
\caption{Fano factor vs. conductance for 119 different Pt atomic
chain configurations. The bold red curve shows the minimum noise
curve when spin degeneracy is imposed. Relaxing spin degeneracy
results in a minimum noise curve shown by the thin black curve.
The inset illustrates the principle of the break junction
experiment.} \label{fig:Fano-vs-G}
\end{figure}

We have recorded similar plots for over 500 configurations of Pt
atomic chains of various length, for which we took seven bias current
points between 0 and 0.44~$\rm{\mu}$A. When the scatter in the plot
of the reduced excess noise was larger than 3\%, or the thermal
noise at start and end of the measurement differed by more than
2\%, we rejected the data. The scatter is mostly due to a large
$1/f$ component in the noise spectrum and the contribution of the
residual amplifier noise correlations to the spectra. After this
selection 119 configurations remain. Figure~\ref{fig:Fano-vs-G}
shows the Fano factors determined from these 119 sets of shot
noise measurements.

The bold red curve shows the minimum noise curve when spin
degeneracy is imposed. Relaxing spin degeneracy results in a
minimum noise curve shown by the thin black curve. The blue broken
lines show the {\em maximum} noise that can be obtained with $N=4$
or $N=6$ (spin) channels. This maximum is obtained by taking all
channels to have the same transmission probability $\tau =
Gh/e^2N$, leading to $F=1-\tau$. The measured data points form a
diffuse cloud in $(G,F)$-space, with its center of mass near
$G=1.5~(2e^2/h)$ and a spread in the conductance in agreement
with the position and width of the first peak in the conductance
histogram. A large fraction of the points lie above the line
labeled $N=4$, which represents the maximum Fano factor when only
four channels are available. This shows that these Pt atomic
chains have at least five conductance channels, in agreement with
calculations \cite{nielsen02,fernandez05, pauly06, smogunov08}. The points
below the blue broken line $N=4$ can be explained by four
channels, but for the majority of points the only conclusion we
can draw is that {\em at least} four channels are involved.

The most striking observation, 
is that all Fano factors for the Pt chain configurations
fall on, or above, the curve describing the minimum noise for
{\em spin-degenerate} channels.  More than 15\% of the measured
points are even found to coincide within the error bars with the
minimum-Fano curve for spin-degenerate channels, and none of the
points are found significantly below it. For spin-split
conductance channels the limiting curve is represented by the thin
black curve in Fig.~\ref{fig:Fano-vs-G} \cite{roche04,dicarlo06}. 
We found no points falling between the spin-degenerate and the spin polarized limit curves.
This suggest that all the conductance channels in the Pt
atomic chains formed in the experiment are effectively spin degenerate, implying at first 
sight that the Pt chain contacts, at variance with results from 
density functional theory (DFT) calculations
\cite{delin03,fernandez05,smogunov08,smogunov08a,thiess09}, are nonmagnetic, as we have argued previously \cite{preprint}.
However, we will investigate this result further below by an analysis based on model calculations, from which we learn 
that the observations are still consistent with a magnetic state for the Pt atomic chain.

Most points in Fig.~\ref{fig:Fano-vs-G} are measured for Pt chains of three
to four atoms in length, and only occasionally five or six atoms. We do not find
any systematic evolution of the Fano factor with stretching of the
chain.  This is instructive, since a large low-temperature spin polarization is believed to be
inevitable in strongly strained Pt chains \cite{smogunov08a}. 
While increasing the length of the chain in steps the Fano 
factor may be seen to touch the minimum noise curve, but 
then it jumps away again from the curve to higher Fano factor values at next steps (see Ref.~\cite{suppl} Fig.~S3). 
A further remarkable observation is the fact that there is a large group
of 18 points that coincide with the curve describing the minimum
noise power for {\it two spin-degenerate channels}. 
These are observations that we would like to explain, for which we now turn to modeling of our system.  

\section{Theoretical modeling}

\begin{figure}[t!]
\includegraphics[width=9cm]{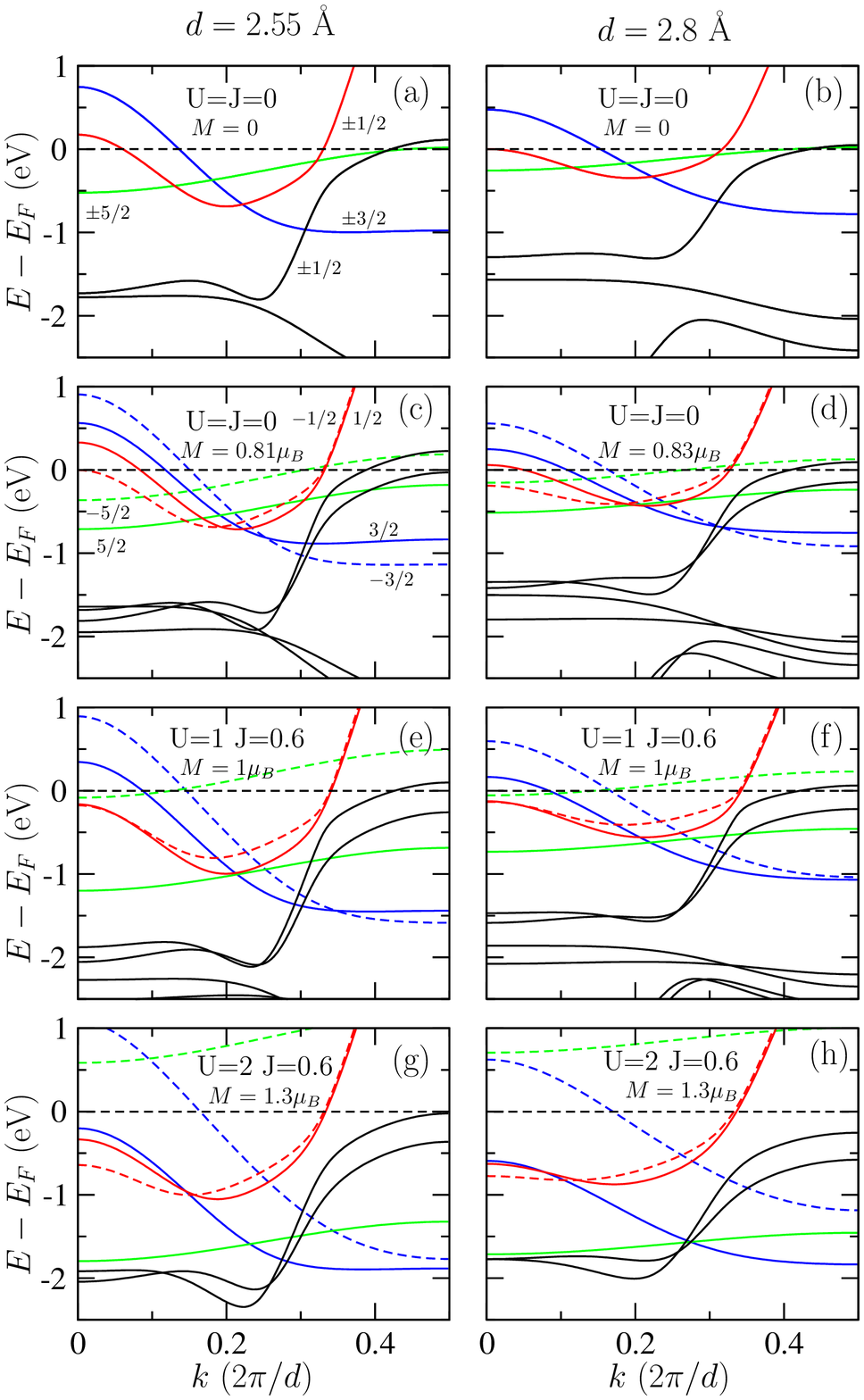}
\caption{Tight-binding relativistic band structures of infinite Pt chains at interatomic
spacings of $d=2.55$ \AA~ (low strain, shown in the left column) and $2.80$ \AA (large strain, right column).
(a)-(d) for $U=J=0$; (e,f) for $U=1$eV, $J=0.6$eV; (g,h) for $U=2$ eV, $J=0.6$ eV.
For $U=J=0$ the bands of non-spin polarized calculation are shown in (a),(b).
A magnetic state with the spin moment $M$ parallel to the chain axis (c),(d) is lower in energy than a nonmagnetic one in all cases.
Local interactions introduced via $U$ and $J$ only enhance magnetism further.
The various bands important for magnetism and conductance are drawn with different colors (see text)
and are labeled by half-integer total angular momentum $m_j$ (along the chain axis) 
in the upper panels (a--d).
Spin-polarized bands for magnetic solutions with positive (negative) $m_j$ are displayed as full (broken) curves.}   
\label{fig_bands}
\end{figure}

We wish to evaluate the impact of  spontaneous spin polarization of atomic Pt chains
on their electron transmission, and from that their shot noise. 
This requires calculation of transmission and shot noise in a variety of Pt chain
nanocontact configurations, for which we need to address the effect of stretching
as well as local Coulomb interactions. 
For the latter, generally required to remove the self-interaction errors of DFT in narrow band problems,
we expect that it is sufficient to 
remain on the level of a mean-field DFT+$U$+$J$ approach \cite{wierzbowska05}.  
We can anticipate that the effect of adding an on-site repulsion $U$ and
exchange interaction $J$, besides favoring magnetism \cite{wierzbowska05}, 
will be to push some conduction channels away from
the Fermi level and to decrease thus the conductance towards the experimental value of $1.5 G_0$.
The question is then what is the level of spin polarization of the remaining channels crossed by Fermi level,  
and what effect would that have on shot noise.

We restrict our considerations to voltages near zero, so that linear response theory, together
with equilibrium electronic structure calculations, is adequate.  
We will ignore many-body effects such as those leading to zero-bias Kondo conductance anomalies. 
If long Pt chain contacts are magnetic, as theory predicts, 
their overall spin should generally be large, $S \gg \frac{1}{2}$.  
In this case, due to spin-orbit coupling, only the two low-energy states $\pm S$ 
(with magnetic moment parallel to the chain axis) will be available
at low energy, with all others raised, or even canceled, by the large (or `colossal') magnetic anisotropy \cite{smogunov08a}.  
Consequently, no Kondo screening and no zero-bias anomalies are expected, 
except possibly for $S=\frac{1}{2}$, a situation which might be realized in very short contacts. 
In addition, because the calculated energy barriers between states $S$ and $-S$ of the chain are 
generally quite large \cite{smogunov08a},
we anticipate a blocking  temperature safely above 4~K, so that
thermal fluctuation effects should also be negligible during the typical short electron 
traversal time across the contact.

Spin polarized DFT, including spin orbit 
interactions \cite{smogunov08,smogunov08a} and self-interaction 
corrections for $d$ orbitals~\cite{wierzbowska05}, is therefore believed to constitute a valid approach to electronic 
structure, ballistic conductance, and shot noise of transition metal nanocontacts. 
Landauer's scattering approach is used to compute the ballistic conductance 
from the electron transmission at the Fermi energy, $G=G_0 T(E_F)$, as was done, e.g., 
in a previous study of Pt nanocontacts~\cite{smogunov08} using the {\it ab-initio} package Quantum-ESPRESSO (QE) \cite{QE}.

We first discuss briefly the previous spin polarized DFT results on Pt chain nanocontacts of Ref.~\cite{smogunov08,smogunov08a}.
It was found that spontaneous polarization with an easy axis parallel to the chain
emerges for 3-atom chains and longer, with the polarization increasing with chain length and with strain.
The calculations found a large number of conducting channels, resulting
in a total conductance slightly above 2$G_0$, to be compared with an experimental 
break junction histogram conductance peak around 1.5 $G_0$. 
Upon onset of magnetism a modest drop of the calculated ballistic conductance was found, which can be explained 
by the fact that the main electronic states, with $|m_j|=\frac{5}{2}$, driving 
the magnetism have a small transmission due to their narrow-band $d_{xy}, d_{x^2-y^2}$ 
character and the resulting small group velocity at the Fermi level.
On the other hand, the wide-band states which dominate conductance, of $s-d_{z^2}$ and
of $d_{xz}, d_{yz}$ character, are only moderately spin polarized so that the resulting conductance
is only slightly affected by the magnetism.

Since first-principles DFT calculations are very time-consuming in the presence of spin-orbit 
interaction (which is mandatory to describe correctly the band structure of Pt atomic chains) we choose 
a simpler {\it ab-initio} based tight-binding (TB) approach implemented in our code at CEA \cite{tb_code}.
It is a self-consistent approach with the TB parameters adjusted to reproduce 
properly the {\it ab-initio} electronic structure results; in our case, to reproduce 
correctly the band structure of Pt nanowires calculated with QE package.
    
It is instructive to analyze first the infinite Pt chain band structures, which foreshadow the conductance channels in a long chain nanocontact.
We show in Fig.~\ref{fig_bands} the band structures of strained Pt chains (strain being generally present 
in break junctions)
for two interatomic distances, $d=2.55$\AA\  and $2.8$\AA, to be compared with an unstrained equilibrium chain spacing of about $2.35$\AA.
The value of $d=2.8$\AA\   seems excessive -- we find 
it to be the DFT breaking point for the infinite Pt chain -- but we still include it as it was considered in some previous calculations of Pt nanocontacts \cite{fernandez05,wang11}.
As seen from Figs.~\ref{fig_bands}(c) and \ref{fig_bands}(d) the $m_j=\pm \frac{5}{2}$ band (green lines), of $d_{xy}, d_{x^2-y^2}$ character,
is strongly spin split and lies around the Fermi level, so that it can be considered as the main `actor' responsible 
for spontaneous spin polarization. When crossing $E_F$ electrons of this flat band have a small group velocity and 
will be heavily reflected at the chain-electrode junctions, yielding little contribution to the transmitted current.
Conversely, the other bands are rather `spectators' in the magnetization process -- their spin splitting is small, 
and only driven by the `actor'  band via intra-atomic Hund's rule coupling. There are six bands with high group velocities 
at the Fermi level which are expected to contribute to conductance: (i) two $s$-like $m_j=\pm \frac{1}{2}$ channels 
appearing in the middle of the Brillouin zone (red curves);
(ii) two $m_j=\pm \frac{3}{2}$ channels, mainly of $d_{xz}, d_{yz}$ character (blue curves); and (iii) two other $m_j=\pm \frac{1}{2}$ bands with
$k$-vector close to the origin, also mainly of $d_{xz}, d_{yz}$ character (red curves). Note that there are two additional $m_j=\pm \frac{1}{2}$ channels shown by black curves,
which have a low group velocity and low transmission probability; we will ignore these in the further discussion. 

To highlight the effect of local interactions enacted by Coulomb parameters $U$ and $J$,
in Figs.~\ref{fig_bands}(e) and \ref{fig_bands}(f) and in Figs.~\ref{fig_bands}(g) and \ref{fig_bands}(h) we show calculations for $U=1$ eV, $J=0.6$ eV and for $U=2$ eV, $J=0.6$ eV, respectively.
The value of exchange parameter $J$ was fixed to be close to its atomic value since it is 
not much affected by the environment, while the `Hubbard' $U$  
is usually much screened by conduction electrons and is not known a priori.
The main effect is the shift of the flat $m_j=-\frac{5}{2}$ band upward while the $m_j=\pm \frac{1}{2}$ and $m_j=\pm \frac{3}{2}$ 
bands are pushed down. We also see that the magnetism is growing stronger with inclusion of Coulomb corrections.
\begin{figure}[t!]
\includegraphics[width=8cm]{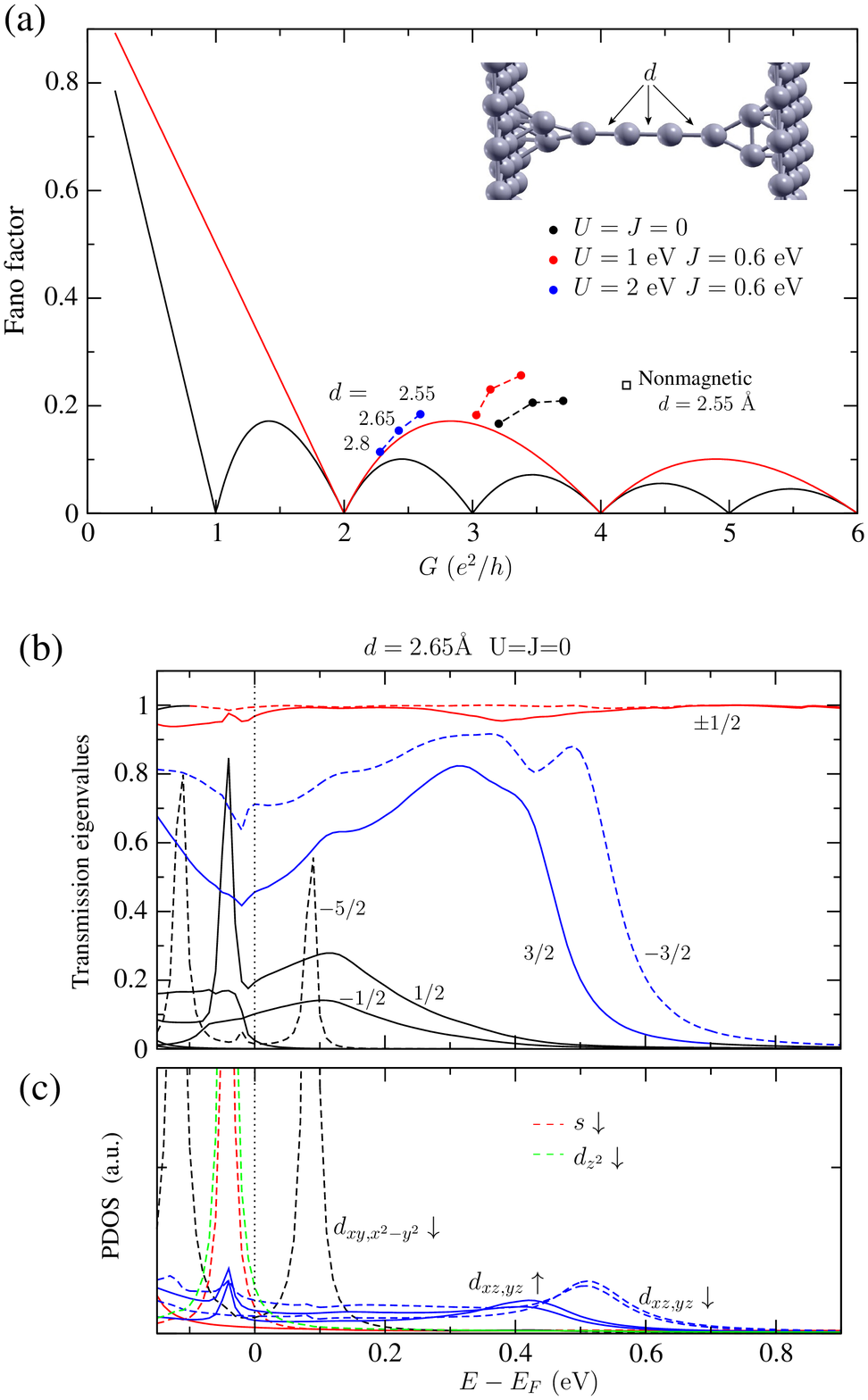}
\caption{Tight-binding results for 4-atom chain Pt nanocontacts (shown in the inset):
(a) Fano factor vs. conductance for $U=J=0$ (black dots), $U=1$ eV, $J=0.6$ eV (red dots),
and $U=2$ eV, $J=0.6$ eV (blue dots). Each group includes the results for three interatomic 
spacings, $d=2.55$\AA, $2.65$\AA, and $2.8$\AA\ and we observe that within each group the conductance 
and the Fano factor decrease
with increasing $d$ (as indicated explicitly for the case of $U=2$ eV, $J=0.6$ eV).  
The nonmagnetic result for $U=J=0$ at $d=2.55$~\AA~ is also shown by the black square. 
(b) Transmission eigenvalues as a function of energy for the same contact, with $U=J=0$ and $d=2.65$~\AA. 
The channels are labeled with their total angular momentum numbers $m_j$ (along the chain axes)
and the position of the Fermi energy is indicated by the vertical dotted line.
(c) The PDOS for the same contact projected on the atomic orbitals of one of the two central 
chain atoms.  
} 
\label{fig_fg}
\end{figure}

In the light of these preliminary considerations, even before a detailed calculation of
transmission and noise, the conduction properties for Pt atomic chain contacts become apparent. The fully spin polarized
`actor' channel will be blocked because of its large mass and weak coupling to the electrodes. 
Only the poorly polarized, less massive `spectator' bands will contribute significantly to conduction.
Their small spin splittings are likely to have a small, possibly even negligible, effect on transmission. 
As a result, conductance is poorly spin sensitive, and so will be shot noise.

To verify these expectations we performed a number of TB calculations 
for four-atom long Pt atomic chains connected to two (111) Pt electrodes by means of two dimers (see the inset in Fig.~\ref{fig_fg}) -- 
a geometry often suggested, see {\it e.g.} Pauly {\it et al.} \cite{pauly06}, for 
Au and Pt atomic chain contacts.
The transmission function in our TB approach is calculated as usual through the non-equilibrium Green's function formula:
$$T(E)=\tr{[\Gamma^L(E)G_c(E)\Gamma^R(E)G_c^+(E)]},$$
where $G_c$ is the retarded Green's function of the central region (including the chain and few
layers of the left and right electrodes) while $\Gamma^L$ and $\Gamma^R$ are coupling matrices of the central region with left and right
electrodes, respectively. To define transmission eigenchannels we follow Jacob {\it et al.} \cite{jacob} -- we choose an atom $i$ in the middle
of the chain and rewrite the total transmission in equivalent form (assuming there is no direct coupling between
the parts left and right of the atom $i$):         
$$T=\tr{[\Gamma_i^LG_i\Gamma_i^RG_i^+]}=\tr{[\Gamma_i^{L1/2}G_i\Gamma_i^RG_i^+\Gamma_i^{L1/2}]},$$     
where now all the matrices refer to the atom $i$ and have their dimension determined by the number of TB orbitals on that atom.
By diagonalizing the last matrix under the trace (which is an Hermitian matrix) one obtains the transmission 
eigenvalues $\tau_n$. The corresponding eigenvectors describe the orbital nature of each eigenchannel at the atom $i$.
  
In Fig.~\ref{fig_fg}(a) we collect all results for the conductance vs. shot noise dependence for
the four-atom chain nanocontact.
There are three groups shown by different colours: for $U=J=0$ (black dots), for $U=1$ eV, $J=0.6$ eV (red dots), 
and for $U=2$ eV, $J=0.6$ eV (blue dots). Each group contains three points, corresponding to different
interatomic spacings in the chain: $d=2.55$, $2.65$, and $2.8$~\AA, and the conductance is always lower for larger interatomic distance $d$,
as indicated explicitly in the case of $U=2$ eV, $J=0.6$ eV.   
The detailed decomposition of the conductances over conductance eigenchannels is given in Table~\ref{table_4atom}.

We analyze first the case of $U=J=0$ (black dots). To obtain more insight into the general nature of conductance and shot noise 
we show in Fig.~\ref{fig_fg}b the transmission eigenvalues as a function of energy for the case of intermediate 
strain, $d=2.65$\AA,  together with the projected density of states (PDOS) for the different atomic orbitals of the central chain atom.
Some general features can be recognized. First, two $m_j = \pm \frac{1}{2}$ chanels, mainly of $s$ character, are almost perfectly
transmitting, almost energy independent, and very poorly spin-polarized 
(strictly speaking, we mark the channels by their {\it main} contribution $m_j$, since the latter is no longer a good quantum number because of broken axial symmetry due to the presence of the electrodes). 
Second, at energies of about $0.6$eV and $0.4$eV above the Fermi level we find, one by one, the appearance
of four channels with $m_j = -\frac{3}{2}$, $\frac{3}{2}$, $\frac{1}{2}$, and $-\frac{1}{2}$, in accordance with the band structure
of the infinite Pt chain (see Fig.~\ref{fig_bands}). These channels are partially spin-polarized and display a density of states (DOS) that is
rather constant in energy, as one may expect for $d_{xz}, d_{yz}$ orbitals. 
We notice also that the $m_j=-\frac{5}{2}$ channel, associated with the `actor' band driving the magnetism,
shows two very narrow peaks in transmission at energies around $-0.1$ and $0.1$~eV, corresponding
with two sharp peaks in the $d_{xy},d_{x^2-y^2}$ DOS. We note in passing that in the nonmagnetic case 
the peak at $0.1$eV is shifted exactly to the Fermi level (and becomes degenerate with the $m_j=\frac{5}{2}$ channel),
which is one of the reasons for the enhanced conductance and shot noise 
calculated 
in the nonmagnetic nanocontact.       
    
We would like now to look at the effect of local Coulomb corrections, simulated by $U$ and $J$, on
the conductance channels and on the shot noise. The band structures for the infinite chain at $U=1$ eV, $J=0.6$ eV (see Figs.~\ref{fig_bands}(e) and \ref{fig_bands}(f))
suggest the closing of the two $m_j= \pm \frac{1}{2}$ channels close to the $\Gamma$ point,
since the corresponding bands are pushed completely below the Fermi level.
Ideally, that could leave us with two perfectly conducting $s$-like channels and two partially polarized 
$m_j = \pm \frac{3}{2}$ channels giving rise to shot noise very close to the `nonmagnetic' boundary
in the plots of Fano factor versus conductance.
In reality, however, the two $m_j= \pm \frac{1}{2}$ channels contribute non-negligible tunneling conductance 
(see Table~\ref{table_4atom}) which keeps the points (red dots in Fig.~\ref{fig_fg}a)
well above the `nonmagnetic' boundary.  

Remarkably, the results for $U=2$ eV, $J=0.6$ eV (blue dots in Fig.~\ref{fig_fg}(a)) produce points that are all very close to the 'non-magentic boundary.'
Here, the $m_j=-\frac{5}{2}$ derived states on the chain become completely emptied
(in line with the empty $m_j=-\frac{5}{2}$ bands in Fig.~\ref{fig_bands}) and are, therefore, irrelevant for electron transport.
The  $m_j= \pm \frac{1}{2}$ channels, largely $s$-like, still transmit almost perfectly for both spin directions, while 
the other pair of $m_j= \pm \frac{1}{2}$ channels are pushed further down from the Fermi level and contribute now very little to the
conductance and noise. The remaining conductance comes from two $m_j = \pm \frac{3}{2}$ channels.
The degree of their spin polarization is, however, not enough to lower the shot noise below the `nonmagnetic' boundary curve.

\begin{center}
\begin{table*}[ht]
\begin{tabular}{ r | *{10}{c|} r | }
\multicolumn{1}{c}{} & \multicolumn{4}{|c|}{$U=J=0$} & \multicolumn{3}{|c|}{$U=1;J=0.6$} & \multicolumn{3}{|c|}{$U=2;J=0.6$} \\ \cline{2-11}
\multicolumn{1}{c|}{$m_j$} & $~2.55~$\AA\  & $~2.55~$\AA\  & $~2.65~$\AA\  & $~2.80~$\AA\  & $~2.55~$\AA\  & $~2.65~$\AA\  & $~2.80~$\AA\  & $~2.55~$\AA\  & $~2.65~$\AA\  & $~2.80~$\AA\   \\ \hline
$-\frac{1}{2}$ & 1.00 & 1.00 & 0.99 & 1.00 & 1.00 & 1.00 & 1.00 & 0.99 & 0.99 & 0.99 \\ \hline
$ \frac{1}{2}$ & 1.00 & 0.99 & 0.98 & 0.98 & 0.98 & 0.98 & 0.98 & 0.98 & 0.99 & 0.99 \\ \hline \hline
$-\frac{3}{2}$ & 0.70 & 0.74 & 0.74 & 0.78 & 0.58 & 0.59 & 0.69 & 0.36 & 0.29 & 0.22 \\ \hline 
$ \frac{3}{2}$ & 0.70 & 0.55 & 0.46 & 0.31 & 0.40 & 0.28 & 0.20 & 0.16 & 0.10 & 0.06 \\ \hline \hline
$-\frac{1}{2}$ & 0.31 & 0.25 & 0.19 & 0.10 & 0.15 & 0.11 & 0.09 & 0.05 & 0.03 & 0.01 \\ \hline 
$ \frac{1}{2}$ & 0.31 & 0.17 & 0.10 & 0.04 & 0.12 & 0.08 & 0.05 & 0.03 & 0.02 & 0.01 \\ \hline \hline
$-\frac{5}{2}$ & 0.11 &      &      &      & 0.15 & 0.09 & 0.02 &      &      &      \\ \hline 
$ \frac{5}{2}$ & 0.11 &      &      &      &      &      &      &      &      &      \\ \hline
\end{tabular}
\caption{Transmission eigenvalues at the Fermi energy for a 4-atom Pt chain contact for 
different $U$, $J$, and interatomic spacings $d$. The first column for $U=J=0$, at a separation of 2.55\AA, shows the data for a non-magnetic calculation. The various channels are labeled by the magnetic quantum number $m_j$.}
\label{table_4atom}
\end{table*}
\end{center}

Extrapolating the results provided by this example, we can draw our main overall lesson. Long 
chain Pt contacts possess three types of channels, $|m_j| = \frac{1}{2}, \frac{3}{2}, \frac{5}{2}$. The first conducts nearly
perfectly and contributes nearly zero noise; the last hardly conducts and also contributes nearly zero
noise; only the second conducts reasonably, and is the main contributor to noise, which is not
much different for a magnetic or a nonmagnetic contact. The amount of conduction in the $|m_j| = \frac{3}{2}$
channel is thus essentially the only quantity that depends on structure and geometry of the break junction. 
Depending  on that, G versus F will give rise to points that accumulate on the `nonmagnetic' boundary -- 
exactly as is observed for the 18 point set -- because that boundary is given by one perfectly transparent
channel plus one partly transparent.  

This comes unexpected at first. Whether spin degenerate or spin polarized, 
first principles DFT calculations predict that, out of a total of 12 atomic ($s$+$d$) states, six to nine spin channels 
should be involved. \cite{nielsen02,vega04,fernandez05,smogunov08}. 
However, DFT calculations usually treat for simplicity highly symmetric idealized
contacts. The more empirical tight binding electronic calculations by Pauly {\it et al.} \cite{pauly06} for example
consider more realistic, low symmetry model geometries, 
and often show a revealing effective reduction of conducting channels down to four,  
particularly in long chain contacts. 
Two of these channels are $s$-orbital dominated, and have a transmission $\tau \simeq 1$, whereas the remaining two 
involve $d$ orbitals and have $\tau \simeq 0.5$. All other channels, also of $d$ character, appear 
to be largely shut off and blocked at these realistic contact conditions.  Therefore, because idealized 
high symmetry contacts appear to be the exception and low symmetry ones the rule, we 
have focused our theoretical  analysis of magnetism on conductance and noise
in the latter case. 

Only when we extend the calculations to longer atomic chains, which are presently inaccessible in experiment, 
and take $U=2$ eV, $J=0.6$ eV, 
we find atomic chain configurations below the nonmagnetic boundary, due to a more pronounced 
spin splitting of the $|m_j| = \frac{3}{2}$
channel. Possibly, future experiments can demonstrate this.

\section{Conclusion}
In conclusion, shot noise for Pt break junctions has been measured and compared with calculations.
Clear evidence for magnetic order in the conductance channels could be obtained, in principle, from shot noise measurements
if one would observe Fano factors below the `non-magnetic boundary'. The experimental observation
of data points that appear to be confined to the region above this boundary at first sight appears to suggest a non-magnetic state. 
The large number of data points that are found {\em on}  the boundary strengthens this impression. 
However, taking realistic atomic chain configurations into account, using spin polarized DFT and TB calculations
employing corrections for e-e interactions (at the level of DFT+$U$+$J$) we find that
conductance and noise involve channels that 
are only weakly affected by magnetism.
The electron transmission properties of Pt atomic chains can be captured by a simple three-channel model, 
with one fully conducting channel, a nonconducting one, and one of intermediate
conductance. The latter dominates the noise signal while the nonconducting one is the main `actor' in driving the magnetism.

\acknowledgments{This work is part of the research programme of
the Foundation for Fundamental Research on Matter (FOM), which is
financially supported by the Netherlands Organisation for
Scientific Research (NWO). 
The calculations at CEA were performed using HPC resources from GENCI-CCRT, 
Grant 2012-x2013096813 on Curie supercomputer.
Work in Trieste was partly sponsored by PRIN/COFIN 2010LLKJBX 004, Sinergia
CRSII2136287/1 and by advances of ERC Advanced Grant 320796 { MODPHYSFRICT. }

\end{document}